\documentclass[aps,prd,preprint,superscriptaddress,showpacs,preprintnumbers,amsmath,amssymb, unsortedadress]{revtex4-1}

\usepackage[utf8]{inputenc}
\usepackage{booktabs}
\usepackage{graphicx}
\usepackage{epstopdf}
\usepackage{dcolumn}
\usepackage{float}
\usepackage[normalem]{ulem}
\usepackage{color}
\usepackage{dcolumn}
\usepackage{multirow}
\usepackage{subfigure}
\usepackage{cancel}
\usepackage{comment}
\usepackage{hyperref}
\usepackage[normalem]{ulem}
\newcommand{\be}{\begin{equation}}
\newcommand{\ee}{\end{equation}}
\newcommand{\bea}{\begin{eqnarray}}

\newcommand{\eea}{\end{eqnarray}}

\newcommand{\pup}{p^\uparrow}

\def\lsim{\mathrel{\rlap{\lower4pt\hbox{\hskip1pt$\sim$}}\raise1pt\hbox{$<$}}}
\def\gsim{\mathrel{\rlap{\lower4pt\hbox{\hskip1pt$\sim$}}\raise1pt\hbox{$>$}}}

\begin{document}

\title{Gluon Sivers function and transverse single spin asymmetries in $e + p^\uparrow \rightarrow \gamma + X$}

\author{Siddhesh Padval}
\email{siddhesh.padval@physics.mu.ac.in}
\affiliation{Department of Physics, University of Mumbai, Mumbai 400098, India}

\author{Rohini M. Godbole}
\email{rohini@iisc.ac.in}
\affiliation{Centre for High Energy Physics, Indian Institute of Science, Bangalore 560012, India}

\author{Abhiram Kaushik}
\email{abhiramkb@imsc.res.in}
\affiliation{The Institute of Mathematical Sciences, HBNI, Taramani, Chennai 600113, India}

\author{Anuradha Misra}
\email{misra@physics.mu.ac.in}
\affiliation{Department of Physics, University of Mumbai, Mumbai 400098, India
}

\author{Vaibhav S. Rawoot}
\email{vaibhavrawoot@gmail.com}
\affiliation{Amity University, Bhatan Post - Somathne, Panvel, Mumbai,Maharashtra 410206, India}

\date{\today}

\begin{abstract}
We present estimates of transverse single-spin asymmetry in prompt photon production in the scattering of low virtuality photons off a polarized proton target and discuss the possibility of using this as a probe to get information about the gluon Sivers function (GSF). Using a generalized parton model (GPM) framework, we estimate the asymmetries at electron-ion collider energy ($\sqrt{s}$ =140 GeV) taking into account both direct and resolved photon processes and find that the dominant contribution,  up to  $10\%$,  comes from  quark Sivers function (QSF) while the contribution  from GSF is found to be up to  $2\%$. However, upon taking into account the effects of the process-dependent initial and final state interactions through the color-gauge invariant generalized parton model approach we find that the situation is significantly changed, with near zero contributions from the QSFs and up to a $1\%$ level contribution from the \textit{f}-type GSF.
Our results indicate that this process may be useful for distinguishing between  GPM and color-gauge invariant generalized parton models and can be used  as  a good probe of \textit{f}-type GSF.

\end{abstract}
 

\maketitle

\section{\label{intro}Introduction}
Transverse single-spin asymmetries (TSSAs) and the underlying physics is an area of hadron physics that has been explored with keen interest  in the past years as these asymmetries provide useful tools for probing the three dimensional and spin structure of the nucleon. Ever since the first observation of large TSSAs in pion production in hadronic collisions involving protons and transversely polarized protons \cite{Klem:1976ui, Adams:1991rw}, a number of experimental and theoretical studies have been performed to measure and understand the TSSAs in various processes \cite{DAlesio:2007bjf,Barone:2010zz}. At present, there are two approaches which are being used to explain these effects, among which one is based on collinear factorization at next-to-leading twist (twist-three) where SSAs are given by convolutions of hard scattering amplitudes with universal quark-gluon-quark and three-gluon correlation functions \cite{Efremov:1984ip,Qiu:1991pp,Qiu:1991wg,Kouvaris:2006zy,Kanazawa:2014dca}. The second approach is based on a generalization of collinear factorization of QCD wherein  the collinear parton distribution functions (PDFs) and fragmentation functions (FFs) are replaced by corresponding transverse momentum dependent (TMD) distribution functions which, on being integrated upon the transverse momentum variables, yield the collinear PDFs and FFs \cite{Collins:2011zzd}. However, such a factorization (TMD factorization) has been established only for two scale processes like semi-inclusive deep inelastic scattering (SIDIS) and Drell-Yan (DY) processes and no such proof exists for single scale processes. A phenomenological approach which uses this for processes even where the TMD factorization has not been established is called the generalized parton model (GPM).  In GPM, TMDs such as Sivers function  are assumed to be process independent. The GPM approach has been used by several groups for estimating asymmetries in processes like $p^\uparrow p \rightarrow \pi X$,  $p^\uparrow p \rightarrow \gamma X$ and open and closed charm production under this assumption\cite{DAlesio:2004eso, Anselmino:2004pnk, Anselmino:2012rq, Godbole:2012bx, Godbole:2016tvq, Godbole:2017syo,Godbole:2017fab}.

The transverse momentum dependent PDFs and FFs are collectively called TMDs and extensive experimental and theoretical work has been done and experiments proposed with the aim of determining these. One of the most  important TMDs is Sivers function which quantifies the azimuthal asymmetry in the distribution of unpolarized quarks and gluons inside a proton which is transversely polarized with respect to the direction of motion\cite{Sivers:1989cc, Sivers:1990fh}. Sivers effect has been used to explain the asymmetries in the process $p^\uparrow + p \rightarrow \pi + X$ \cite{Anselmino:1994tv} and has since then been used in computation of  asymmetries in processes involving a transversely polarized proton target. While the quark Sivers functions (QSFs) have been studied extensively and a number of parametrizations have been proposed based on fits to data \cite{Anselmino:2005ea, Anselmino:2008sga}, not much information is available on gluon Sivers function(GSF). The first indirect estimates  of GSF were obtained  based on GPM in Ref.~\cite{DAlesio:2015fwo} by fitting  the gluon Sivers function to midrapidity data on SSA in $\pi^0$ production at relativistic heavy ion collider (RHIC) and using the QSFs fitted earlier to SIDIS data. However, there is still need to have more direct and clean probes of GSF in which the contribution from QSFs and other TMDs is absent or negligible. Open and closed charm production in $p^\uparrow p$ and low virtuality $p^\uparrow e$ collisions as well as prompt photon production in $p^\uparrow p$ collisions have recently been proposed as probes of GSF within GPM framework\cite{Anselmino:2004pnk,  Godbole:2017fab, Godbole:2014tha,Godbole:2017syo, DAlesio:2017rzj}.
As mentioned earlier, GPM assumes universality of Sivers function  in addition to the assumption of TMD factorization. However, it is now well established that the universality of Sivers function does not hold in general and that the Sivers function is process dependent. 
This process dependence of the Sivers function arises due to the process dependent initial state interactions (ISIs) and final state interactions (FSIs) between the active parton and the spectator partons of the polarized hadron. This observation led to a modification of GPM approach in which the process dependence of Sivers function is taken into account by a careful analysis of initial and final state interactions in one-gluon exchange approximation\cite{Gamberg:2012iq}. The modified GPM approach, now known as color-gauge invariant generalized parton model (CGI-GPM), amounts to absorbing the effects of ISIs and FSIs into the hard part of the process under consideration, thus restoring the universality of the Sivers function. The hadronic cross sections are then obtained by convoluting the universal Sivers function with the modified hard part, which now contains the process dependence and which is obtained by combining the partonic cross section with the initial/final state interactions\cite{DAlesio:2011kkm, DAlesio:2013cfy}. 

CGI-GPM approach has been applied to $J/\psi $ and \textit{D}-meson production at RHIC \cite{DAlesio:2017rzj} and has recently also been applied to the study of TSSA in prompt photon production at RHIC by us and also by D'Alesio {\textit{et al.}} independently \cite{Godbole:2019pp,DAlesio:2019pp}. Both of these studies indicate that prompt photon production in  $p^\uparrow$ $p$  collisions can be used to discriminate between GPM and CGI-GPM approaches. 

In the present work, we have considered TSSAs in the low-virtuality electroproduction of prompt photons at electron-ion collider (EIC) energy using both GPM and CGI-GPM approaches. We have taken into account along with the direct photon contribution, also the resolved photon contribution to prompt photon production where the partonic scattering takes place between a parton originating  from the quasivirtual photon emitted by the incoming electron and the parton from the polarized proton \citep{Godbole:1994, Godbole:1995,Godbole:1998zi}. 
Further in CGI-GPM computation, the major contribution comes from \textit{f}-type GSF while contribution of QSF and \textit{d}-type GSF to asymmetry are nearly zero. Our results indicate that this process can be used to discriminate between GPM and CGI-GPM and can also be used to extract information about the \textit{f}-type GSF. 

The plan of the paper is as follows: in Sec.~\ref{formalism}, we give expressions for the differential cross section and TSSAs. In Sec.~\ref{formalism2}, we discuss the CGI-GPM approach and give  expressions for the modified hard parts for the relevant processes. In Sec.~\ref{tmds}, parametrizations used for the gluon Sivers function and the quark Sivers function are given. Finally, in Sec.~\ref{results}, we present  numerical estimates  of TSSAs in both GPM and CGI-GPM frameworks.

\section{\label{formalism}Prompt Photon Production in the GPM formalism}

\begin{figure}[h]
\begin{center}
\includegraphics[width=7cm, height=5cm]{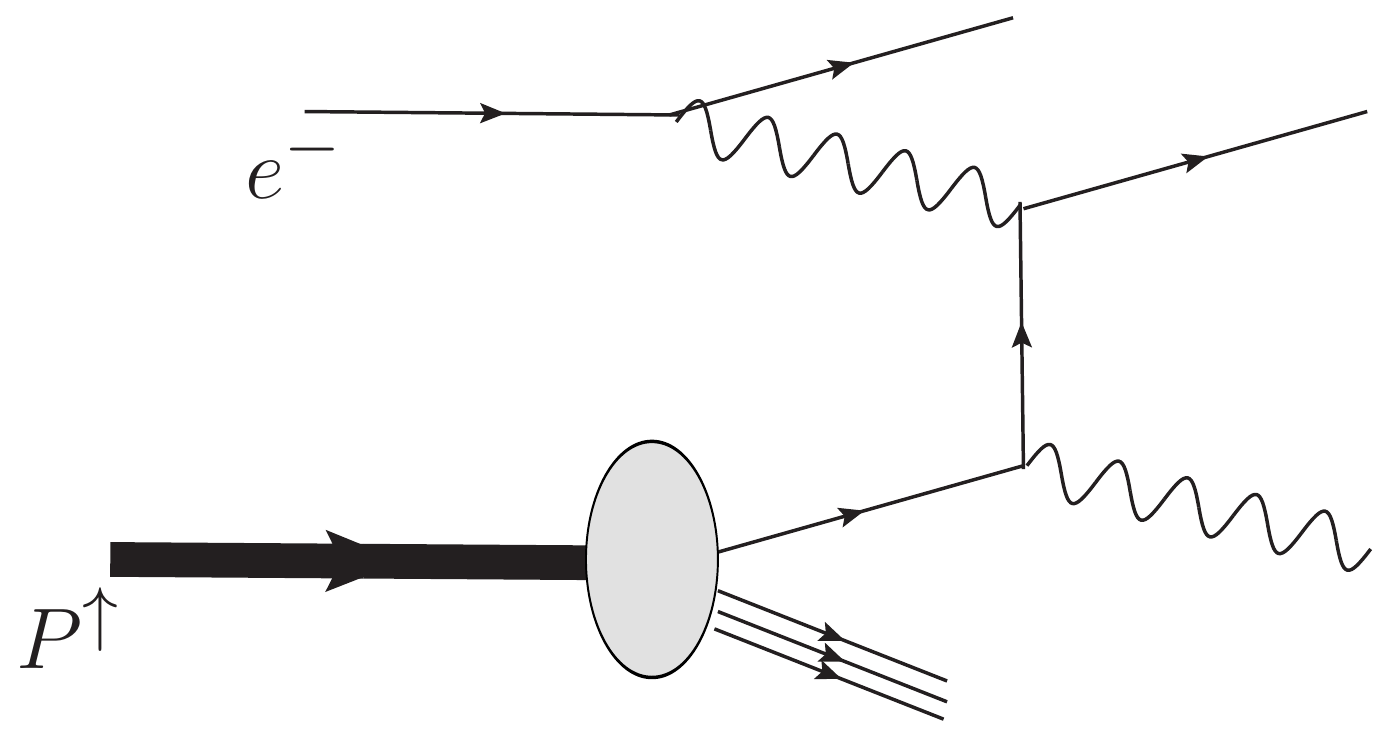}
\hspace*{1cm}
\includegraphics[width=7cm, height=5cm]{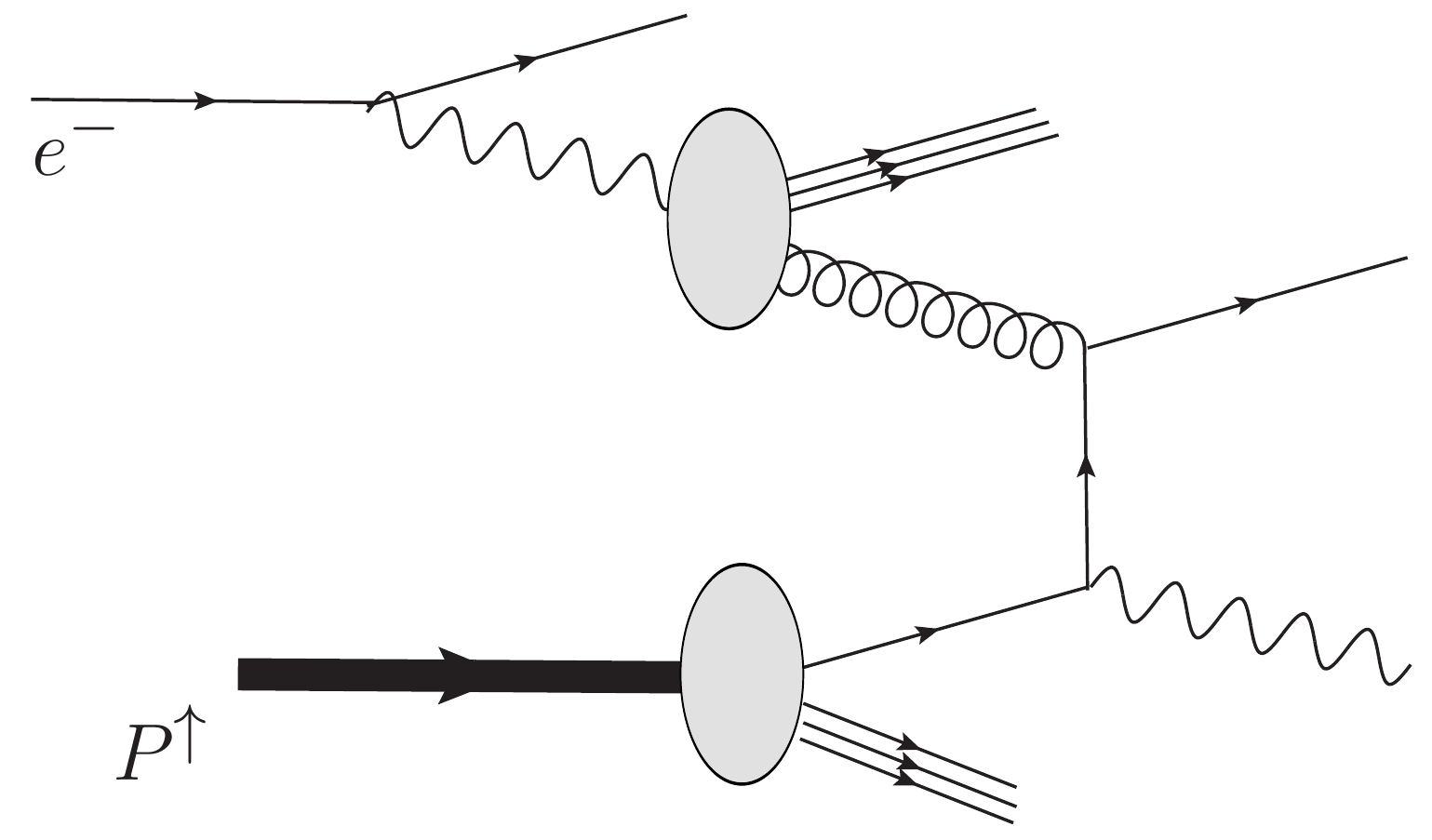}
\caption{Representative diagrams for prompt photon production through direct subprocess (left) and through the resolved subprocess
(right), in lepton-proton collisions. We consider a proton moving in the $+Z$ direction, with a polarization along the $+Y$ axis. The lepton
is unpolarized moving along the $Z$ direction.}
\label{fig:repre_diags}
\end{center}
\end{figure}

We consider low-virtuality electroproduction of prompt photons in which the scattered lepton is unobserved since it is deflected at a very small angle thus making the photon emitted by the initial electron almost real. This can be seen from the fact that the four-momentum squared of the incoming photon can be approximated by $p_{\gamma_I}^{2} = 2EE^{\prime}(1 - cos(\theta))$ with $E$ and $E^{\prime}$ being the energy of the initial and final electron, respectively.
Prompt photon production in electron proton collisions can take place through direct electromagnetic process $\gamma + p^\uparrow \rightarrow \gamma + X $ or through a resolved process in which a parton from the photon participates in the hard process. An illustration of both mechanisms of prompt photon production is given in Fig.\ref{fig:repre_diags}. The partonic subprocess in the direct contribution, $q\gamma\to\gamma+X$ is $\mathcal{O}(\alpha^2)$, whereas the parton level processes involved in the resolved contribution, $q\bar{q}\to\gamma g$, $qg\to\gamma q$ and $gq\to\gamma q$ are $\mathcal{O}(\alpha\alpha_{s})$. Even though the resolved subprocesses are of a higher order in $\alpha_{s}$ as compared to  the direct process, the parton distribution in the photon has a leading behavior proportional to $\alpha/\alpha_{s}$. Hence the resolved processes also contribute to the hadronic cross section effectively  at $\mathcal{O}(\alpha^2) $. In addition, there is also a fragmentation contribution of $\mathcal{O}(\alpha^2)$, wherein the  final state photon originates from the fragmentation of a final state quark or gluon of the partonic subprocess. This  contribution can be removed by putting isolation cuts in the experiments and therefore we have not included the fragmentation  contribution in this work. We have also not taken into account  $\gamma g \rightarrow \gamma g$ process as it contributes at $\mathcal{O}(\alpha^{2}_s \alpha^2 $) and is therefore not relevant for our discussion at  leading order (LO). 

In the following, we consider the process $e+p\to\gamma+X$ with the proton moving in the $+z$ direction and the electron moving in the $-z$ direction with the final state photon produced in the $x$-$z$ plane. In this work, we are interested in the transverse single-spin asymmetry which is given by
\be
A_N=\frac{d\sigma^\uparrow-d\sigma^\downarrow}{d\sigma^\uparrow+d\sigma^\downarrow}
\label{Eq:SSA}
\ee
where $d\sigma^{\uparrow( \downarrow)}$ is the invariant differential cross section with the proton polarized in the $\uparrow$ ($\downarrow$) direction, i.e., in the $+y$ ($-y$) direction. The denominator in Eq.~(\ref{Eq:SSA}) is twice the unpolarized cross section. In the GPM approach, the contribution of the direct process to the denominator in Eq.~(\ref{Eq:SSA}) is given by

\begin{eqnarray}
	(d\sigma^\uparrow+d\sigma^\downarrow)_{Direct} &=&  E_{\gamma}\frac{d\sigma^{e p^{\uparrow} \rightarrow \gamma X}}{d^3\textbf{P}_{\gamma}} + E_{\gamma}\frac{d\sigma^{e p^{\downarrow} \rightarrow \gamma X}}{d^3\textbf{P}_{\gamma}}\\ \nonumber &=& 2 \sum_{a=q,\bar{q}} \int dx_{\gamma_{I}}dx_{a}d^{2}\textbf{k}_{\perp a}
	f_{\gamma/e}(x_{\gamma_{I}})\hat{f}_{a/p}(x_a,\textbf{k}_{\perp a})
	\frac{\hat{s}}{x_{\gamma_{I}}x_{a}s}\frac{\hat{s}}{\pi}\frac{d\sigma^{a \gamma \rightarrow \gamma d}}{dt}
	\delta(\hat{s}+\hat{t}+\hat{u})\\ \nonumber
	&=& 2 \sum_{a=q,\bar{q}}\int dx_{\gamma_{I}}dx_{a}d^{2}\textbf{k}_{\perp a}
	f_{\gamma/e}(x_{\gamma_{I}})\hat{f}_{a/p}(x_a,\textbf{k}_{\perp a})
	\frac{\alpha^{2}}{x_{\gamma_{I}}x_{a}s}
	H^{U}_{a \gamma \rightarrow \gamma d}
	\delta(\hat{s}+\hat{t}+\hat{u}).
\end{eqnarray}
In the above expression $x_{a(\gamma_I)}$ is the light cone momentum fraction of the incoming parton (photon) along the parent proton (electron) direction, $\textbf{k}_{\perp q}$ is the intrinsic transverse momentum of the parton with respect to the proton's direction, and the Mandelstam variables are defined as $\hat{s} = (p_a + p_{\gamma_I})^2$, $\hat{t} = (p_a - p_{\gamma})^2$ and $\hat{u} = (p_{\gamma_I} - p_{\gamma})^2$, where $p_a$ is the momentum of the parton, $p_{\gamma_I}$ is the momentum of the initial state photon from the electron, and $p_\gamma$ is the momentum of the final state electron. The details of the kinematics are presented in the Appendix.

Here, $\hat{f}_{q/p}(x_q,\textbf{k}_{\perp q})$ is the unpolarized transverse momentum dependent parton distribution function and $f_{\gamma/e}(x_{\gamma_{I}})$ is the William-Weizacker distribution of quasireal photons in electron which is given by,
\begin{eqnarray}  \label{photonflux}
f_{\gamma/e}(x_{\gamma_{I}})=\frac{\alpha}{2\pi }\left[
2m_e^2x_{\gamma_{I}} \left(\frac{1}{Q^2_{\mathrm{min}}}-\frac{1}{Q^2_{\mathrm{max}}}\right)+\frac{1+(1-x_{\gamma_{I}})^2}{x_{\gamma_{I}}}\ln\frac{Q^2_{\mathrm{max}}}{Q^2_{\mathrm{min}}}\right ]
\end{eqnarray}
where $\alpha$ is the electromagnetic coupling and $Q^2_{\mathrm{min}}=m_e^2\frac{x_\gamma^2}{1-x_\gamma}$ with $m_e$ 
being the electron mass. $Q^2_{\mathrm{max}}$ corresponds to the maximum virtuality of the exchanged photon. Since we are interested in a photoproduction process with an exchanged photon of low virtuality and small momentum transfer, we set $Q^2_{\mathrm{max}}=1$ motivated by the COMPASS antitagging cuts.

At leading order in $\alpha_{s}$, the direct process is sensitive to only the quark content of the proton through the $q\gamma\to\gamma q$ process. The hard part for it is given by
\begin{equation}
	H^{U}_{q \gamma \rightarrow \gamma q}=-2e_{q}^{4}\bigg( \frac{\hat{t}}{\hat{s}}+\frac{\hat{s}}{\hat{t}} \bigg).
\end{equation}

The contribution of the direct process to the numerator of Eq.~(\ref{Eq:SSA}) is given by
\begin{eqnarray}\label{Eq:num_dir}
(d\sigma^\uparrow-d\sigma^\downarrow)_{Direct}  &=&  E_{\gamma}\frac{d\sigma^{e p^{\uparrow} \rightarrow \gamma X}}{d^3\textbf{p}_{\gamma}} - E_{\gamma}\frac{d\sigma^{e p^{\downarrow} \rightarrow \gamma X}}{d^3\textbf{p}_{\gamma}}\\ \nonumber
&=& \int dx_{\gamma_{I}} dx_{q}d^{2}\textbf{k}_{\perp q}
f_{\gamma/e}(x_{\gamma_{I}}) \Delta^{N}{f}_{q/p^{\uparrow}}(x_q,\textbf{k}_{\perp q})
\frac{\alpha^{2}}{x_{\gamma_{I}}x_{q}s}
H^{U}_{q \gamma \rightarrow \gamma q}
\delta(\hat{s}+\hat{t}+\hat{u}).
\end{eqnarray}
In the above expression, $\Delta^Nf_{q/p^\uparrow}(x,\textbf{k}_{\perp q})$ is the Sivers function for the quark $q$, which describes an azimuthal asymmetry in the transverse momentum distribution of unpolarized quarks inside a transversely polarized proton,
\begin{equation}
\begin{split}
\Delta^{N}{f}_{a/p^{\uparrow}}(x_a,\textbf{k}_{\perp a})
&=\hat{f}_{a/p^{\uparrow}}(x_a,\textbf{k}_{\perp a})-\hat{f}_{a/p^{\uparrow}}(x_a,-\textbf{k}_{\perp a})\\&=\hat{f}_{a/p^{\uparrow}}(x_a,\textbf{k}_{\perp a})-\hat{f}_{a/p^{\downarrow}}(x_a,\textbf{k}_{\perp a})\\
&=\Delta^{N}{f}_{a/p^{\uparrow}}(x_a,k_{\perp a})\cos\phi_{a}\\
&=-2\frac{k_{\perp a}}{M_{p}}f_{1T}^{\perp a}(x_a,k_{\perp a})\cos\phi_{a}.
\end{split}
\end{equation}

The functional forms we use for both the unpolarized TMD  $\hat{f}_{q/p}(x_q,\textbf{k}_{\perp q})$ and the Sivers function are given in Sec.~\ref{tmds}.

The contributions of the resolved process to, respectively, the denominator and numerator of Eq.~(\ref{Eq:SSA}), are given by
\begin{eqnarray}\label{Eq:den_res}
(d\sigma^\uparrow+d\sigma^\downarrow)_{Resolved}  &=&  E_{\gamma}\frac{d\sigma^{e p^{\uparrow} \rightarrow \gamma X}}{d^3\textbf{p}_{\gamma}} + E_{\gamma}\frac{d\sigma^{e p^{\downarrow} \rightarrow \gamma X}}{d^3\textbf{p}_{\gamma}}\\ \nonumber
&=& 2\sum_{a,b=g,q,\bar{q}} \int dx_{\gamma_{I}}dx_{a}dx_{b}
d^{2}\textbf{k}_{\perp a}
f_{\gamma/e}(x_{\gamma_{I}})
f_{b/\gamma}(x_b)
\hat{f}_{a/p}(x_a,\textbf{k}_{\perp a})\\ \nonumber
&& \frac{\alpha\alpha_{s}}{x_{a}x_{b}x_{\gamma_{I}}s}
H^{U}_{a b \rightarrow \gamma d}\delta(\hat{s}+\hat{t}+\hat{u})
\end{eqnarray}
and
\begin{eqnarray}\label{Eq:num_res}
(d\sigma^\uparrow-d\sigma^\downarrow)_{Resolved}  &=&  E_{\gamma}\frac{d\sigma^{e p^{\uparrow} \rightarrow \gamma X}}{d^3\textbf{p}_{\gamma}} - E_{\gamma}\frac{d\sigma^{e p^{\downarrow} \rightarrow \gamma X}}{d^3\textbf{p}_{\gamma}}\\ \nonumber
&=& \sum_{a,b=g,q,\bar{q}} \int dx_{\gamma_{I}}dx_{a}dx_{b}
d^{2}\textbf{k}_{\perp a}
f_{\gamma/e}(x_{\gamma_{I}})
f_{b/\gamma}(x_b)
\Delta^{N}{f}_{a/p^{\uparrow}}(x_a,\textbf{k}_{\perp a})\\ \nonumber
&& \frac{\alpha\alpha_{s}}{x_{a}x_{b}x_{\gamma_{I}}s}
H^{U}_{a b \rightarrow \gamma d}\delta(\hat{s}+\hat{t}+\hat{u}).
\end{eqnarray}
In the above expressions $x_{\gamma_{I}}$, $x_a$ and $x_b$  are the light cone momentum fractions of the incoming photon, incoming parton in the proton, and incoming parton from the photon, respectively. 
$f_{b/\gamma}(x_b)$ is the resolved photon distribution of parton $b$ in the photon.

The hard parts for the partonic processes involved in the  resolved contribution are,
\begin{equation*}
	H^{U}_{q \bar{q} \rightarrow \gamma g}=\frac{N_{c}^{2}-1}{N_{c}^{2}}e_{q}^{2}\bigg( \frac{\hat{u}}{\hat{t}}+\frac{\hat{t}}{\hat{u}} \bigg),\hspace{5mm}
	H^{U}_{q g \rightarrow \gamma q}=-\frac{e_{q}^{2}}{N_{c}}\bigg( \frac{\hat{t}}{\hat{s}}+\frac{\hat{s}}{\hat{t}} \bigg),\hspace{5mm}
	H^{U}_{g q \rightarrow \gamma q}=-\frac{e_{q}^{2}}{N_{c}}\bigg( \frac{\hat{u}}{\hat{s}}+\frac{\hat{s}}{\hat{u}} \bigg).\\
\end{equation*}

In principle, in Eqs.~(\ref{Eq:den_res}) and (\ref{Eq:num_res}) the resolved photon distribution also depends on $k_T$. However it is not an intrinsic transverse momentum since it does not arise out of any internal structure. In this initial study of Sivers asymmetry in $ep\to\gamma+X$, which is related to the transverse motion of partons in the proton, we have neglected the transverse-momentum dependence of parton-in-photon distribution functions.

\section{\label{formalism2}The CGI-GPM formalism}
In the generalized parton model, it is assumed that all TMDs are universal and  therefore, Sivers function extracted in SIDIS can be used to make predictions for inclusive particle production in hadron hadron scattering. However, we know that operator definition of Sivers function contains Wilson lines (gauge links), which are required for the gauge invariance of Sivers function \cite{Collins:2002kn}. This presence of Wilson lines makes Sivers function process dependent. In the CGI-GPM, this process dependence is taken into account by considering the ISI between active parton from incoming unpolarized hadron and the spectators  and  the FSI between active parton from the final state and the spectators of incoming polarized hadron \cite{DAlesio:2017rzj,Gamberg:2012iq,DAlesio:2011kkm,DAlesio:2013cfy}
. The ISIs and FSIs are then approximated by single eikonal gluon exchange, which is equivalent to the leading order expansion of Wilson line in the strong coupling constant $g_{s}$. These ISI/FSIs provide an imaginary part to the amplitude that is required for the existence of a SSA. The only effect of these ISI/FSI is to modify the color factor of the participating processes. In CGI-GPM, this modified color factor is absorbed into the hard part leading to process dependent hard parts in GPM expressions. These modified hard parts now can be used alongwith the universal Sivers function in the calculation of the asymmetry.

\begin{figure}[!h!]
	\begin{center}
		\includegraphics[width=7.5cm, height=4.5cm]{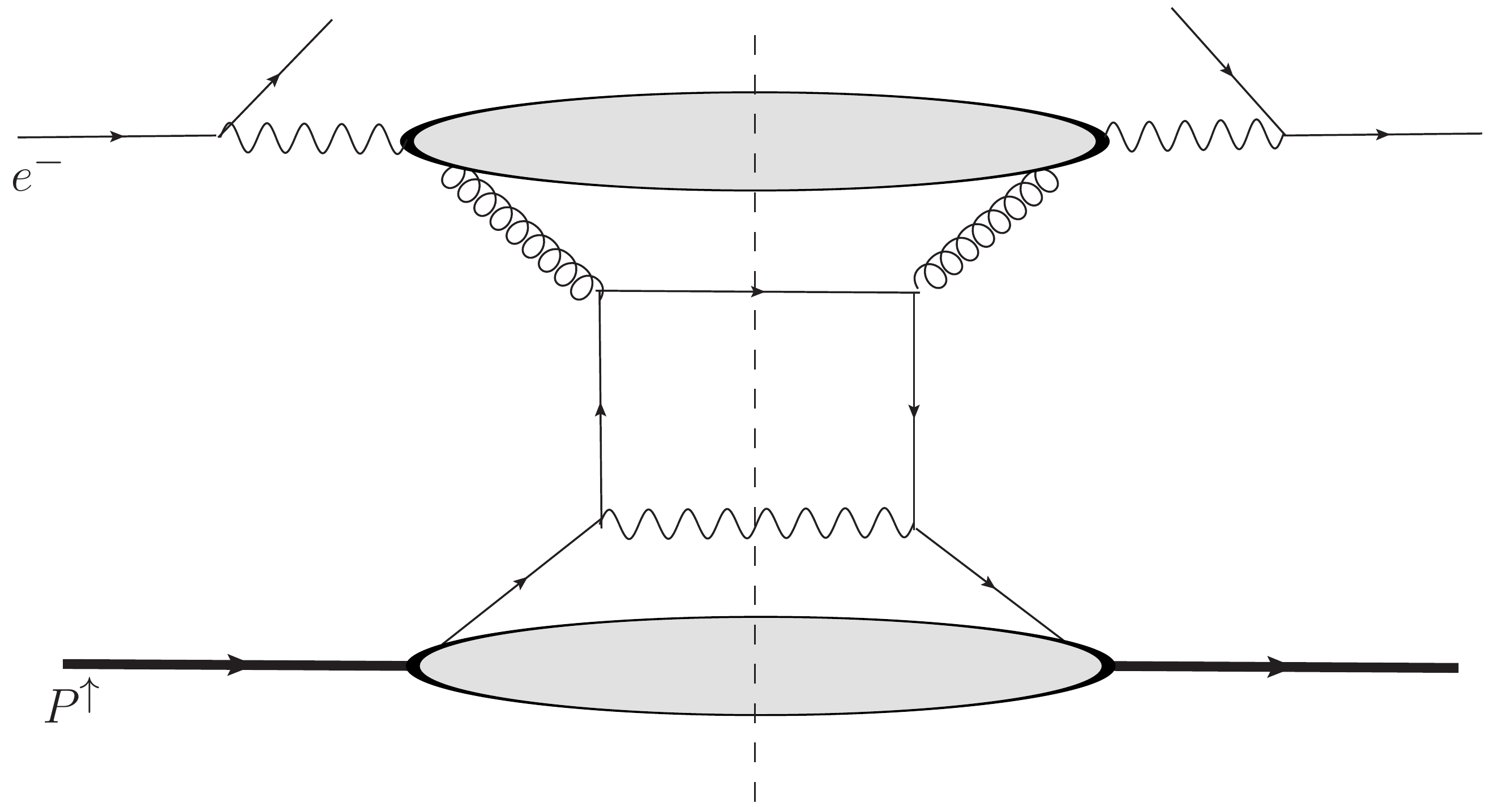}
		\hspace*{1cm}
		\includegraphics[width=7.5cm, height=4.5cm]{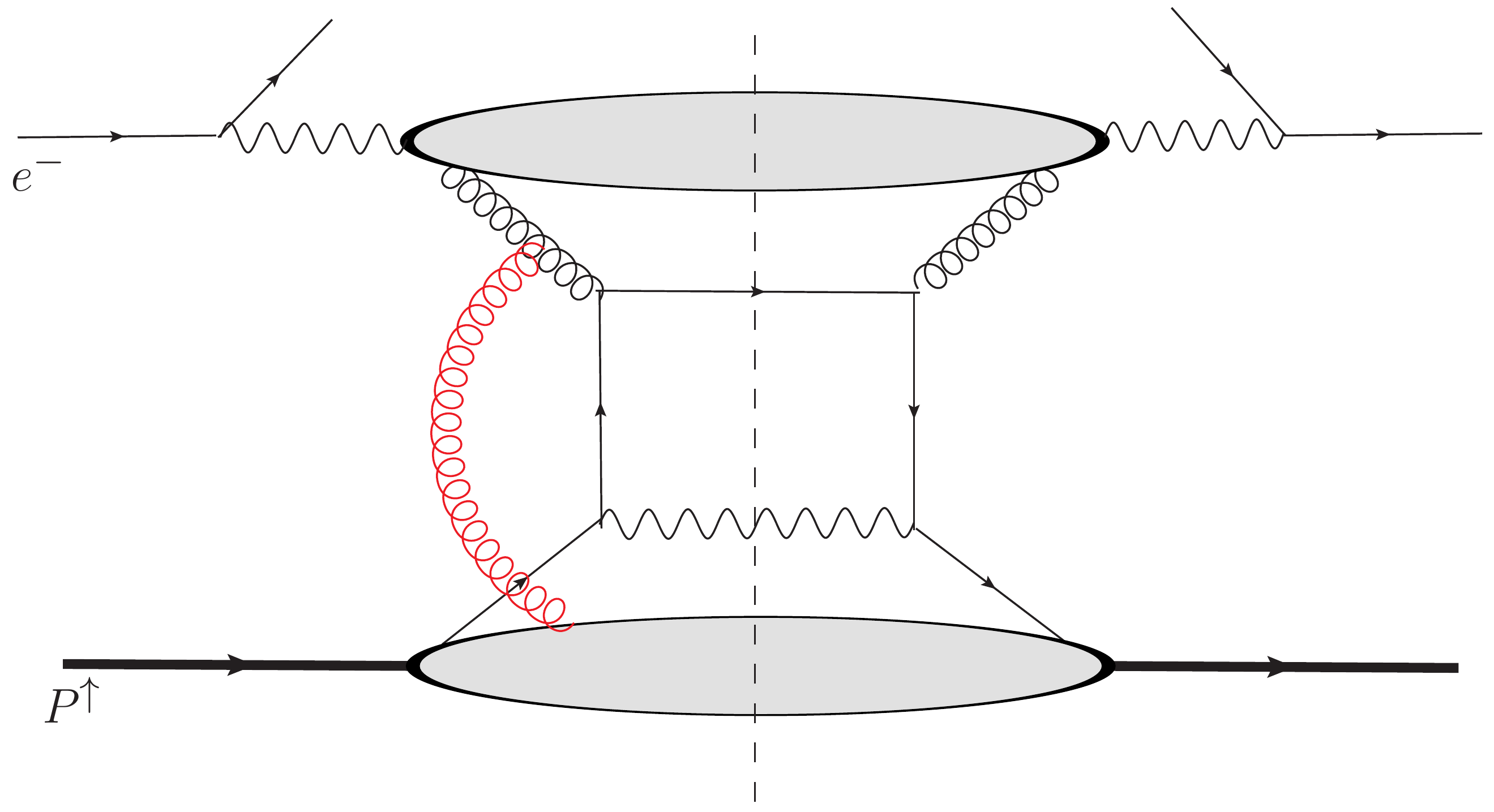}
		\caption{LO diagrams for the resolved subprocess $q g \rightarrow \gamma q$ in GPM formalism (a) and in CGI-GPM formalism(b). Here only ISI contributes through resolved channel.}
		\label{fig:rescgidaigs}
	\end{center}
\end{figure}

To illustrate this, we have shown an example in Fig.\ref{fig:rescgidaigs}. Figure.\ref{fig:rescgidaigs}(a) is  the resolved subprocess $q g \rightarrow \gamma q$ in GPM formalism, and Fig.\ref{fig:rescgidaigs}(b) is the same process with FSI in CGI-GPM formalism. This approach was first proposed in Ref.\cite{Gamberg:2012iq} for the process $ p^{\uparrow} p \rightarrow \pi + X $and has been extended to the case of GSF in Ref.\cite{DAlesio:2017rzj} for  $ p^{\uparrow} p \rightarrow J/\psi + X $ and $ p^{\uparrow} p \rightarrow D + X $. To obtain the asymmetry  in CGI-GPM approach,  one has to make the following replacement for the QSF in Eqs. (\ref{Eq:num_dir}) and (\ref{Eq:num_res}),
\begin{equation}
f_{1T}^{\perp}H^{U}_{q b \rightarrow c d} \rightarrow  f_{1T}^{\perp}H^{mod}_{q b \rightarrow c d}=\frac{C_{I}+C_{Fc}}{C_{U}}f_{1T}^{\perp}H^{U}_{q b \rightarrow c d}
\end{equation}
Here,  $C_{U}$ is the color factor corresponding to unpolarized cross section, $C_{I}$ and $C_{Fc}$ are color factors for the case of ISI and FSI, respectively and $f_{1T}$ is the Sivers function fitted to SIDIS data.

\begin{figure}[!h!]
	\begin{center}
		\includegraphics[width=7.5cm, height=4.5cm]{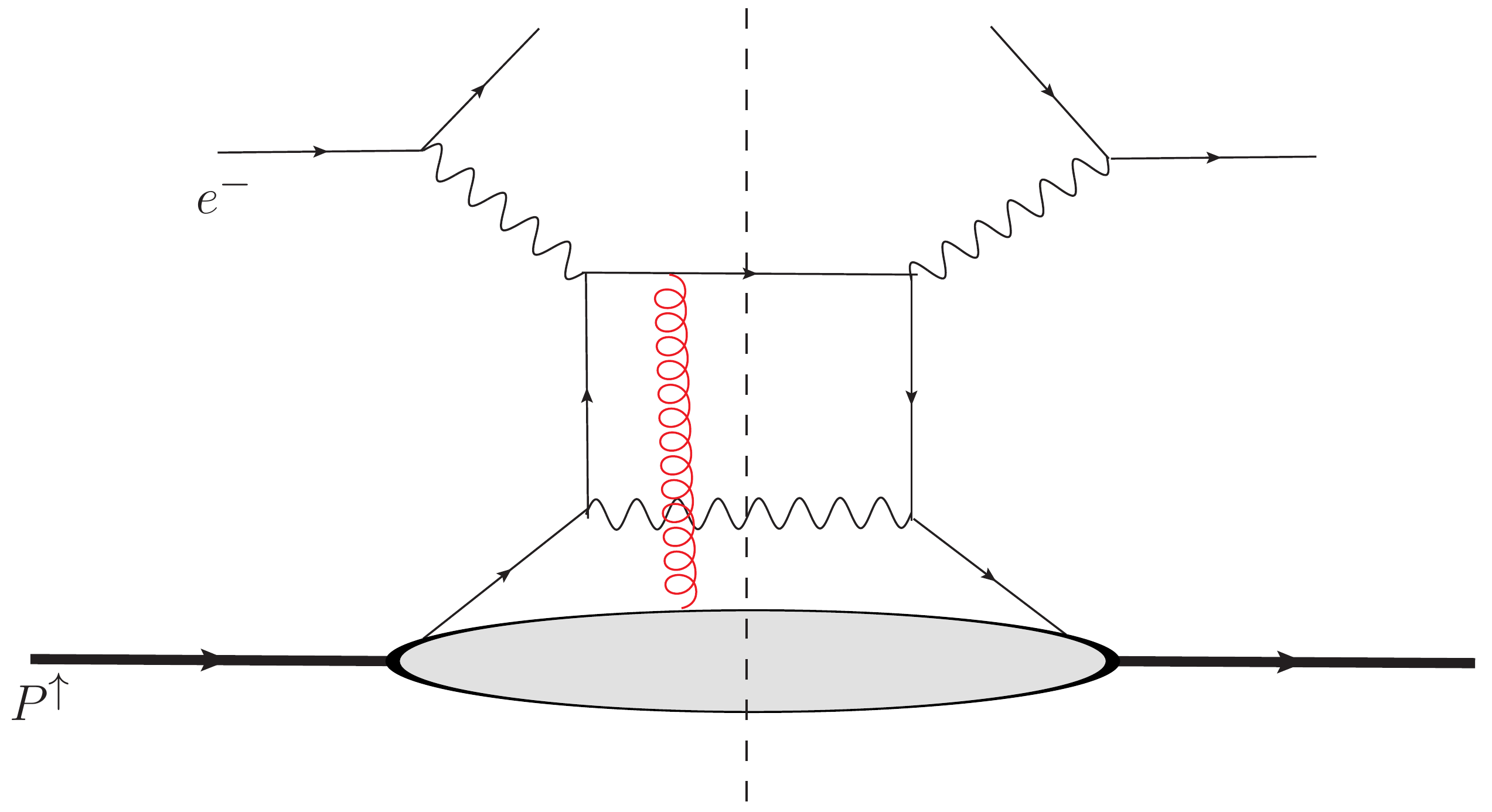}
		\includegraphics[width=7.5cm, height=4.5cm]{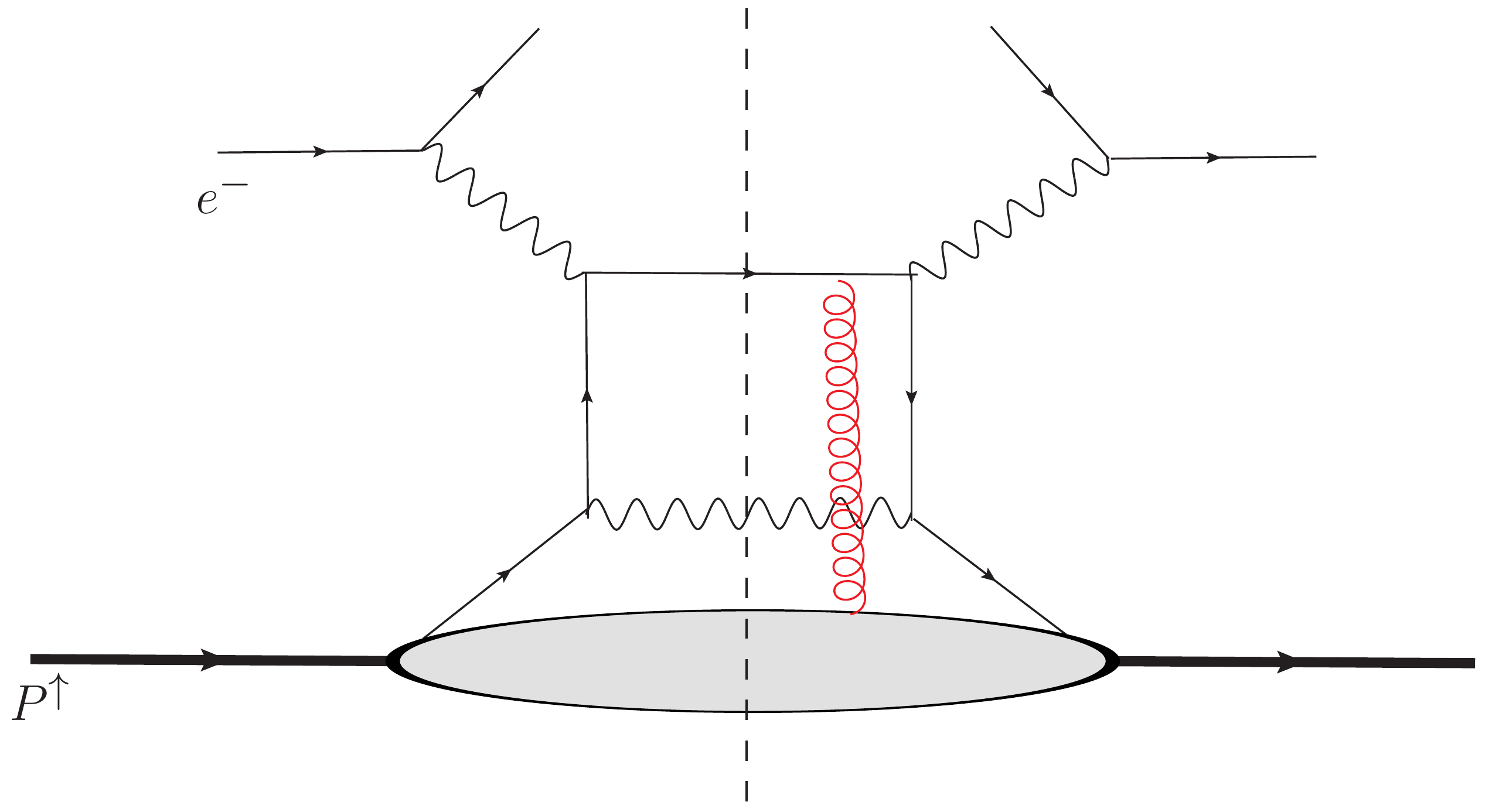}
		\caption{LO diagrams of FSI of an unobserved particle for the direct subprocess $q \gamma \rightarrow \gamma q$ for the unobserved particle. }
		\label{fig:dircgidaigs}
	\end{center}
\end{figure}

In the CGI-GPM framework, the process-dependent gluon
Sivers function can be written as a linear combination of two independent universal gluon distributions $f_{1T}^{\perp g(f)}$ and $f_{1T}^{\perp g(d)}$,  which correspond to the two different ways in which color can be neutralized \cite{DAlesio:2017rzj}. Therefore, for the case of GSF the substitution required in Eq. (\ref{Eq:num_res}) is
\begin{equation}
\begin{split}
f_{1T}^{\perp g}H^{U}_{g b \rightarrow c d} \rightarrow  f_{1T}^{\perp}H^{mod}_{g b \rightarrow c d}
=&\frac{C_{I}^{f}+C_{Fc}^{f}}{C_{U}}f_{1T}^{\perp g(f)}H^{U}_{g b \rightarrow c d} + \frac{C_{I}^{d}+C_{Fc}^{d}}{C_{U}}f_{1T}^{\perp g(d)}H^{U}_{g b \rightarrow c d} \\
=&f_{1T}^{\perp}H^{(f)}_{g b \rightarrow c d}+f_{1T}^{\perp}H^{(d)}_{g b \rightarrow c d}\\
\end{split}
\end{equation}
Now, for the case of direct subprocess $q \gamma \rightarrow \gamma q$, both initial particle and the final observed particles are photons which cannot emit eikonal gluon. Hence, there are no ISI/FSI diagrams to provide an imaginary part in any of the amplitudes that are interfering (which is the case with  the standard unpolarized amplitude), and hence there will not be any SSA. In other words, there is a modified hard-part for direct subprocess $q \gamma \rightarrow \gamma q$, but it vanishes. It should be noted that this argument is valid only when considering the CGI-GPM framework wherein the ISI/FSI is explicitly taken to be a prerequisite for a nonzero SSA. In the context of standard GPM framework where the Sivers function is considered to be universal, the standard hard part does contribute to the numerator of the asymmetry. Finally, we would like to stress the point that although there are no ISI and FSI in  direct subprocess $q \gamma \rightarrow \gamma q$, there is FSI from the unobserved particle which is the quark. However, contribution from this FSI vanishes, when we sum the different cut diagrams which are shown in Fig. \ref{fig:dircgidaigs}

The modified hard parts for the processes under consideration have been calculated in \cite{Godbole:2019pp,DAlesio:2019pp} and are given below for the sake of completeness : 

\begin{equation*}
	H^{mod}_{q \bar{q} \rightarrow \gamma g}=
	-H^{mod}_{\bar{q} q \rightarrow \gamma g}=
	\frac{e_{q}^{2}}{N_{c}^{2}}
	\bigg( \frac{\hat{u}}{\hat{t}}+\frac{\hat{t}}{\hat{u}} \bigg)\\
\end{equation*}
\begin{equation*}
	H^{mod}_{q g \rightarrow \gamma q}=
	-H^{mod}_{\bar{q} g \rightarrow \gamma q}=
	-\frac{N_{c}}{N_{c}^{2}-1}e_{q}^{2}\bigg( \frac{\hat{t}}{\hat{s}}+\frac{\hat{s}}{\hat{t}} \bigg)\\
\end{equation*}
\begin{equation*}
	H^{(f)}_{g q \rightarrow \gamma q}=
	H^{(f)}_{g \bar{q} \rightarrow \gamma \bar{q}}=
	-\frac{1}{2}H^{U}_{g q \rightarrow \gamma q}\\
\end{equation*}
\begin{equation*}
	H^{(d)}_{g q \rightarrow \gamma q}=
	-H^{(d)}_{g \bar{q} \rightarrow \gamma \bar{q}}=
	\frac{1}{2}H^{U}_{g q \rightarrow \gamma q}\\
\end{equation*}

\section{\label{tmds}Parametrization of the TMD}
For the unpolarized TMDs, we adopt the commonly used form with the collinear PDF multiplied by a Gaussian transverse momentum dependence,
\be\label{Eq:paroftmd}
f_{i/p}(x,k_\perp;Q)=f_{i/p}(x,Q)\frac{1}{\pi\langle k_\perp^2\rangle}e^{-k_\perp^2/\langle k_\perp^2\rangle}
\ee
with $\langle k_\perp^2\rangle=0.25\text{ GeV}^2$ \cite{Anselmino:2005nn}.

The Sivers function is generally parametrized as,
\be
\Delta ^N f_{i/\pup}(x,k_{\perp};Q)=2\mathcal{N}_i(x)f_{i/p}(x,Q)\frac{\sqrt{2e}}{\pi}\sqrt{\frac{1-\rho}{\rho}}k_\perp \frac{e^{-k^2_\perp/\rho\langle k^2_\perp\rangle}}{\langle k^2_\perp\rangle^{3/2}}
\ee
with $0<\rho<1$. Here $\mathcal{N}_i(x)$ is a function that parametrizes the $x$ dependence of the Sivers function. For the Sivers function to satisfy the positivity bound,
\be
\frac{|\Delta^Nf_{i/p^\uparrow}(x,\mathbf{k}_\perp)|}{2f_{i/p}(x,\mathbf{k}_\perp)}\leq 1 \hspace*{1cm}\>\forall \>x, \mathbf{k}_\perp,
\label{posbound}
\ee
it is necessary to have  $|\mathcal{N}_i(x)|<1$. A commonly adopted functional form for $\mathcal{N}_i(x)$ that ensures that the positivity bound is satisfied for all values of $x$ is given by,
\be
\mathcal{N}_i(x)=N_ix^{\alpha_i}(1-x)^{\beta_i}\frac{(\alpha_i+\beta_i)^{\alpha_i+\beta_i}}{\alpha_i^{\alpha_i}\beta_i^{\beta_i}}.
\ee

In this work, in order to study the efficacy of the probe, we use Sivers functions with the positivity bound saturated, viz. $\mathcal{N}_i(x)=1$ and $\rho=2/3$. The parameter $\rho$ is set to 2/3 in order to maximize the first $k_\perp$ moment of the Sivers function, following Ref.~\cite{DAlesio:2010sag}. We will refer to this as the ``saturated'' Sivers function. We are using the value of the TMD width obtained by Anselmino \textit{et al}. in Ref.\cite{Anselmino:2005nn} by analyzing the Cahn effect in unpolarized
SIDIS which then has been used for fitting the parameters in quark Sivers function in Refs.\cite{Anselmino:2005ea, Anselmino:2008sga}. Since the asymmetries do depend on the $\langle k_\perp^2\rangle$, for consistency we use the same value while using saturated QSF and GSF as well.

\section{\label{results}NUMERICAL ESTIMATES}
\begin{figure}[h]\begin{center}
\includegraphics[width=8cm, height=7cm]{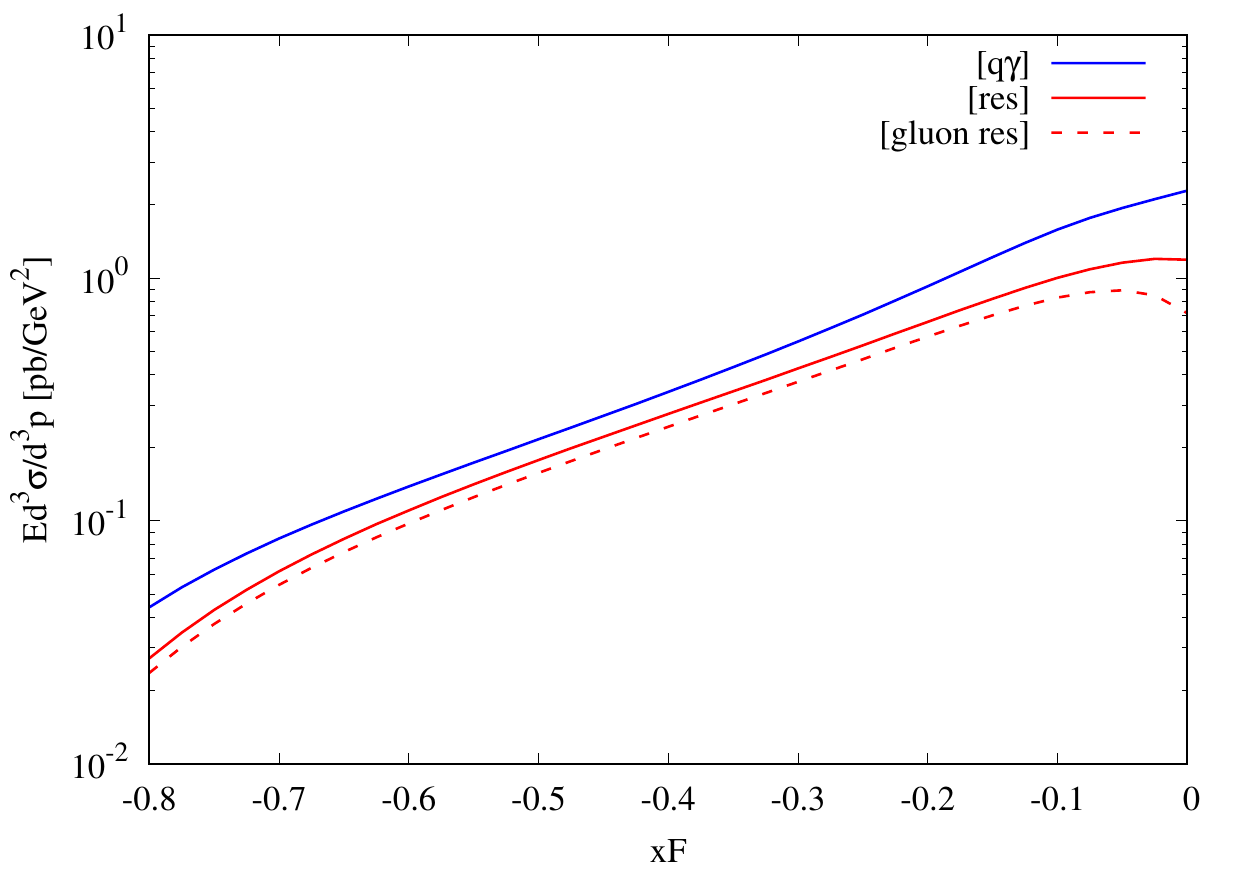}
\includegraphics[width=8cm, height=7cm]{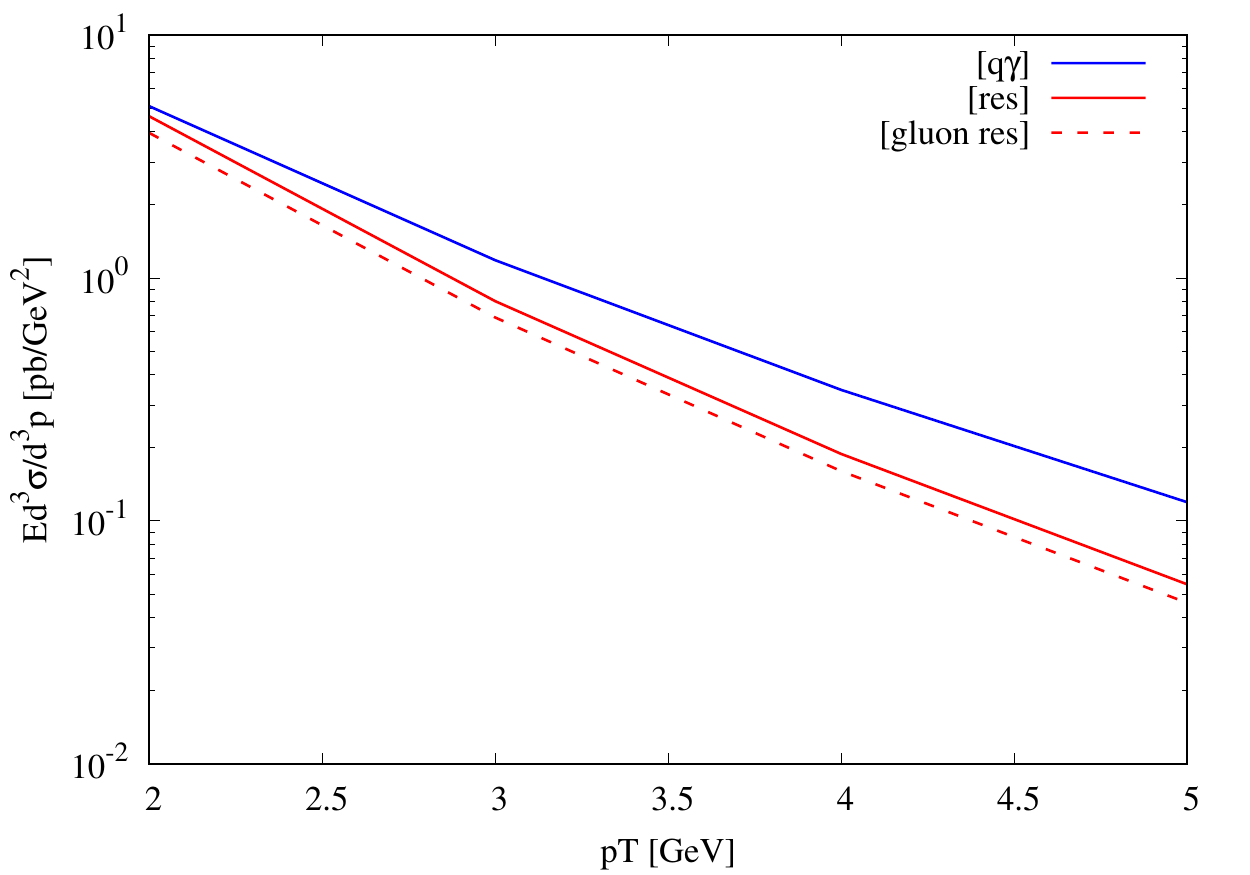}
\caption{Differential cross section $E\frac{d^{3}\sigma}{d^{3}{p}}$ for the inclusive prompt photon production at the EIC energy ($\sqrt{s}=140$ GeV) as  a function of $x_{F} $ ( at $ p_{T} = 3 $ GeV, left panel) and $p_{T}$ (at rapidity $\eta =-2$, right panel). Blue line shows the contribution from the direct subprocess and red line shows the contribution from resolved subprocesses. Dashed red line indicates the contribution from gluon initiated resolved process.}
\label{fig:unpol}
\end{center}
\end{figure}

First, we consider the unpolarized Lorentz-invariant cross section for the production of prompt photons at EIC energy, $\sqrt{s}=140 $ GeV.  In Fig.~\ref{fig:unpol}, we show $x_{F}$ distribution of cross section in the left panel at fixed $p_{T} = 3$ GeV  and $p_{T}$ distribution of cross section in the right panel at fixed rapidity $\eta=-2$.
Fragmentation contributions are not considered here since, as mentioned earlier, these can be eliminated by applying isolation cuts in the experiment. 
We have used CTEQ6L \cite{Pumplin:2002vw} PDFs for the collinear parton distribution functions and the AFG04 \cite{Aurenche:2005} parton distribution in the photon. A flavor independent Gaussian width  $\langle k_\perp^2\rangle=0.25\text{ GeV}^2$ has been used for TMD-PDFs. Factorization scale $Q$ for evaluating the PDFs and $\alpha_{s}$ in the cross section expressions is chosen to be the transverse momentum of produced photon $p_{T}$.
\par As evident from Fig.~\ref{fig:unpol}, the contribution to the unpolarized cross section from direct subprocess $q \gamma \rightarrow \gamma q$ is dominant in comparison to the  resolved contribution. As far as resolved contributions are concerned, the gluon initiated process, i.e., $g q \rightarrow \gamma q$, where gluon is coming from proton and quark is coming from the photon, is the dominant one amongst the resolved subprocesses since the region that we have considered here, which is $x_{F}<0$, where low values of $x_{a}$ are being probed. At these low values of momentum fraction the distribution of gluon inside the proton is dominant over the distribution of quarks. The fact that the cross section for direct subprocess is of the same order as the cross section from total resolved contribution will be crucial for discriminating between GPM and CGI-GPM formalism. This is due to the fact that  the  direct subprocess has no initial state or  final state interaction and therefore, it does not contribute to the numerator of the asymmetry in the CGI-GPM formalism.

\begin{figure}[!ht]
\begin{center}
\includegraphics[width=8cm, height=7cm]{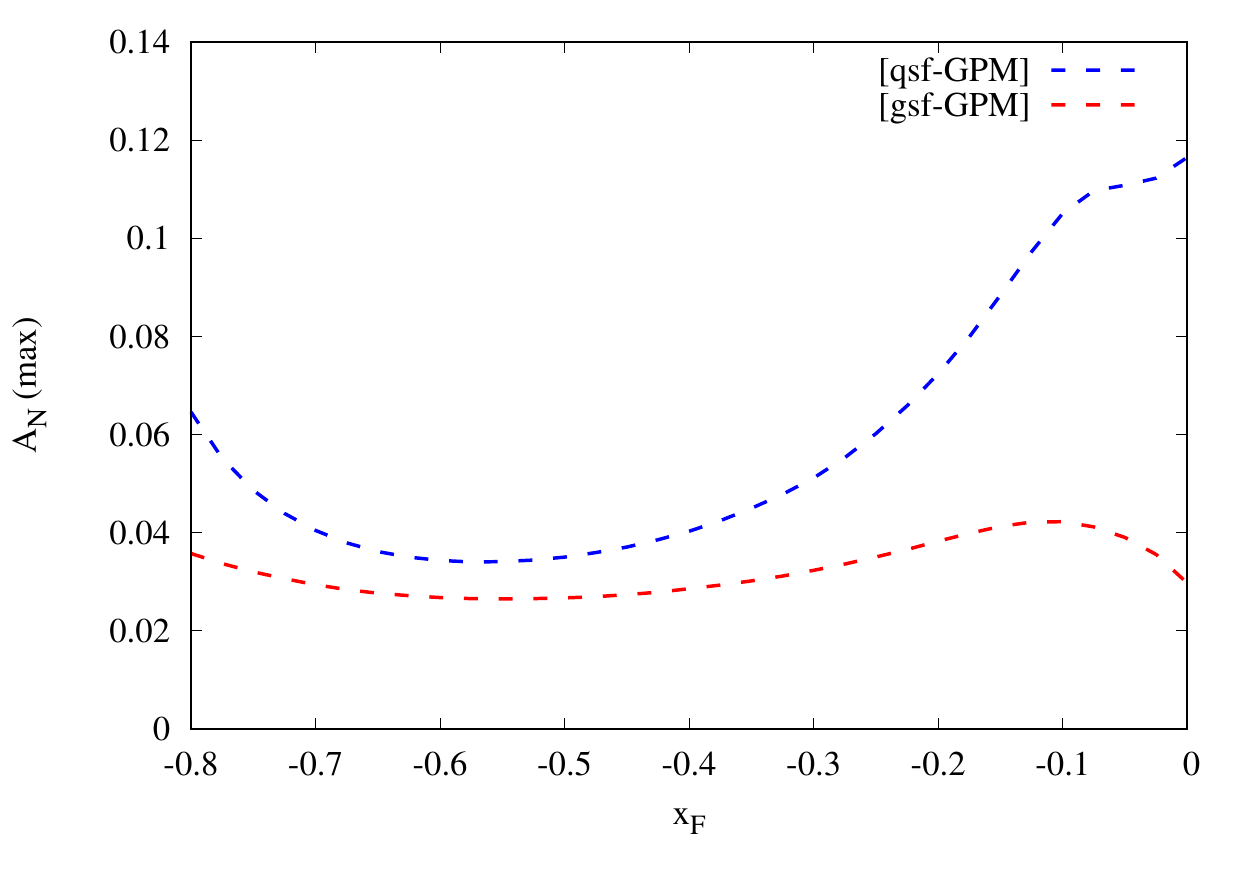}
\includegraphics[width=8cm, height=7cm]{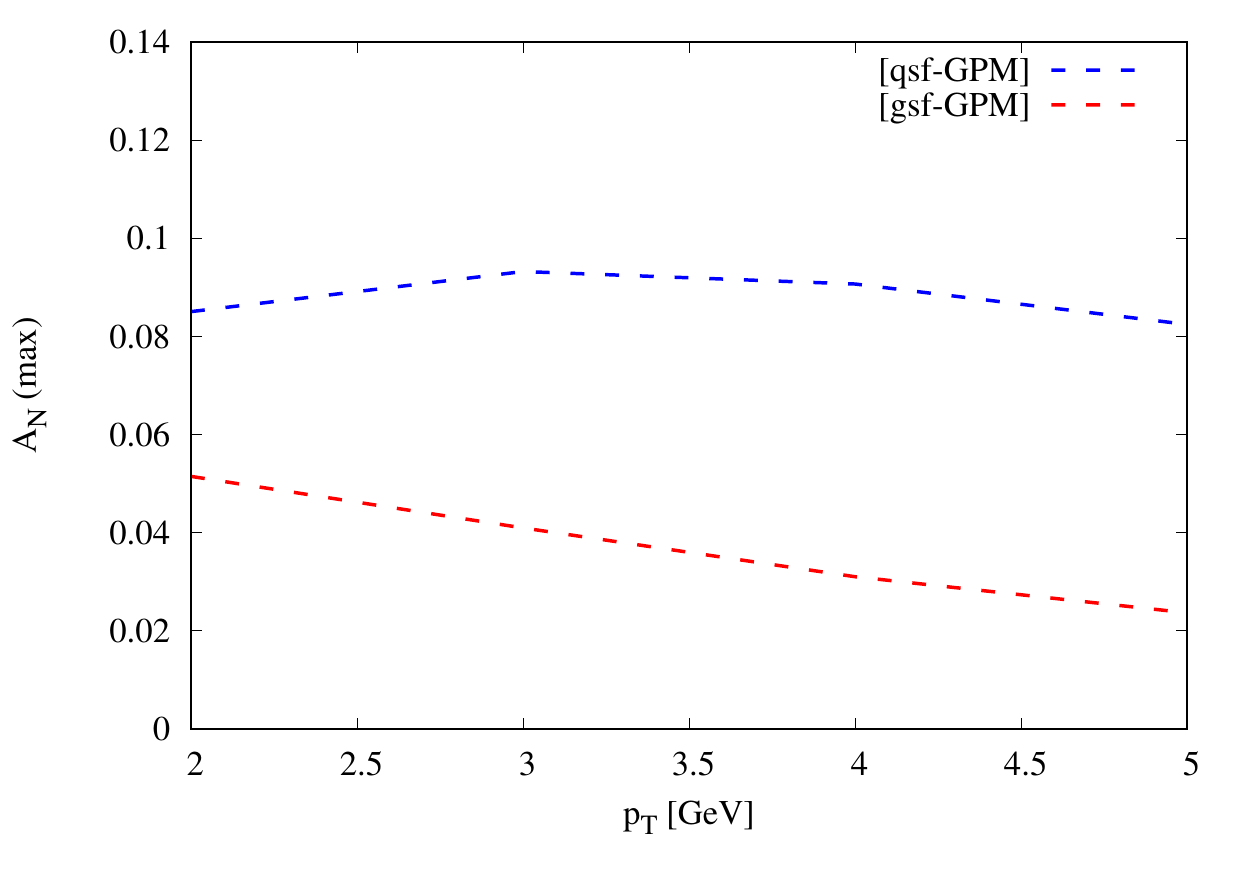}
\caption{ Estimates for single spin asymmetry in prompt photon production using saturated quark and gluon Sivers function in GPM. Dashed blue line indicates contribution from quark Sivers function and dashed red line indicates contribution from gluon Sivers function. Results are obtained using CTEQ6L \cite{Pumplin:2002vw} parametrization for collinear PDFs for partons in the proton and AFG04 \cite{Aurenche:2005} parametrizations for parton distribution in the photon.}
\label{fig:an_gpm}
\end{center}
\end{figure}

Next we consider the asymmetry in the GPM framework. In Fig. \ref{fig:an_gpm}, we show asymmetry estimates using saturated quark and saturated  gluon Sivers functions.
 
As shown in left panel of  Fig.\ref{fig:an_gpm}, asymmetry in GPM is dominated by quark Sivers function. At fixed  $p_{T}$ value of  3 GeV, the saturated QSF contribution  to asymmetry in $x_F$ distribution is found to be up to 13 \% at midrapidity and up to 3\% to 7\% at $x_{F} < -0.2$, whereas saturated gluon Sivers function contribution is  up to 3\% to 4\% in this region. In the right panel of Fig.\ref{fig:an_gpm}, we show the asymmetry estimates for $p_{T}$ distribution at fixed rapidity $\eta = -2$. In this case, we find that saturated QSF contributes 8.5\% to the SSA, whereas saturated GSF contributes up to 5\% at $p_{T} =2$ GeV. 

Finally, we consider the estimates of SSA in the CGI-GPM framework. In this model, we relax the assumption of universality of Sivers function by taking into account  the initial and final state interactions. Note that we do not need to consider the direct subprocess in the numerator of SSA 
for the reason mentioned earlier. However, it does contribute to the denominator which is twice the unpolarized cross section.

\begin{figure}[!ht]
\begin{center}
\includegraphics[width=8cm, height=7cm]{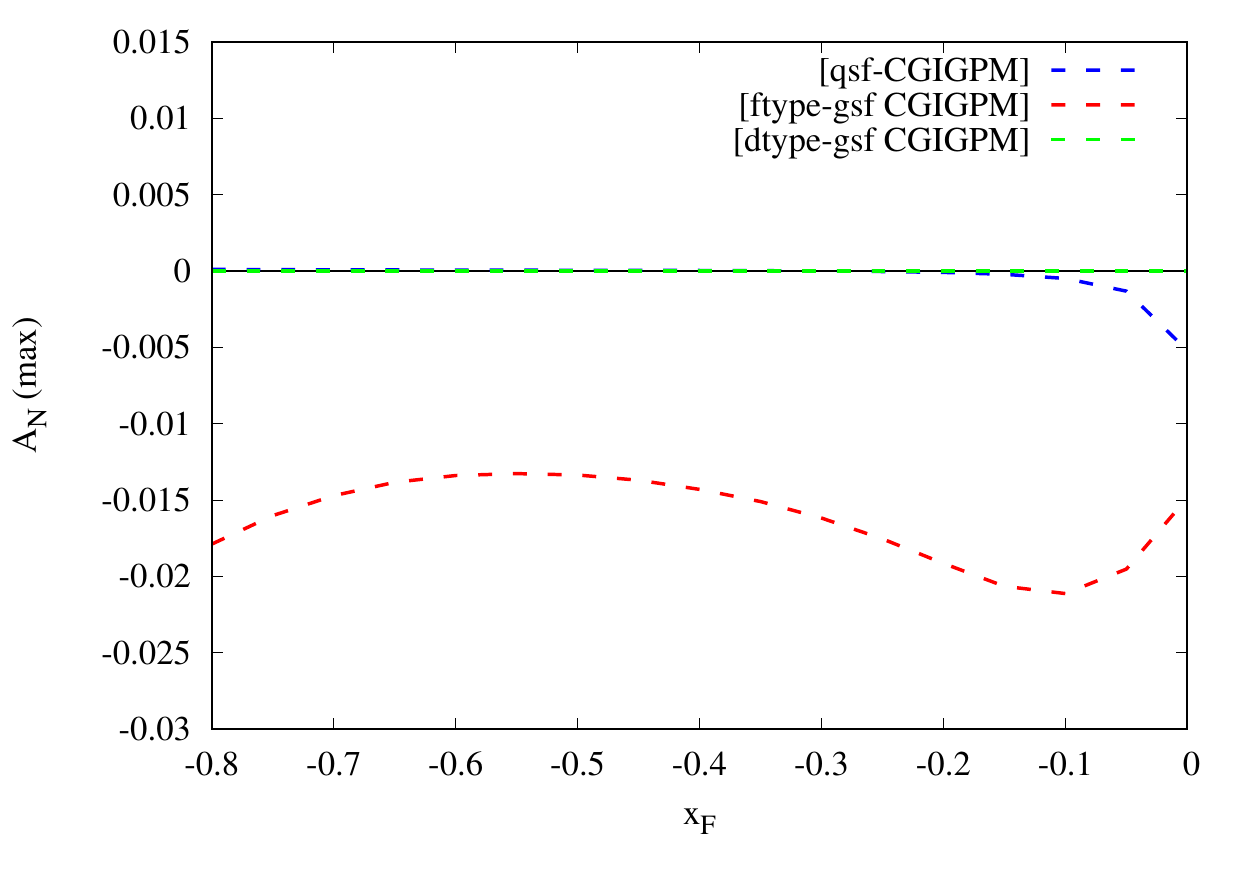}
\includegraphics[width=8cm, height=7cm]{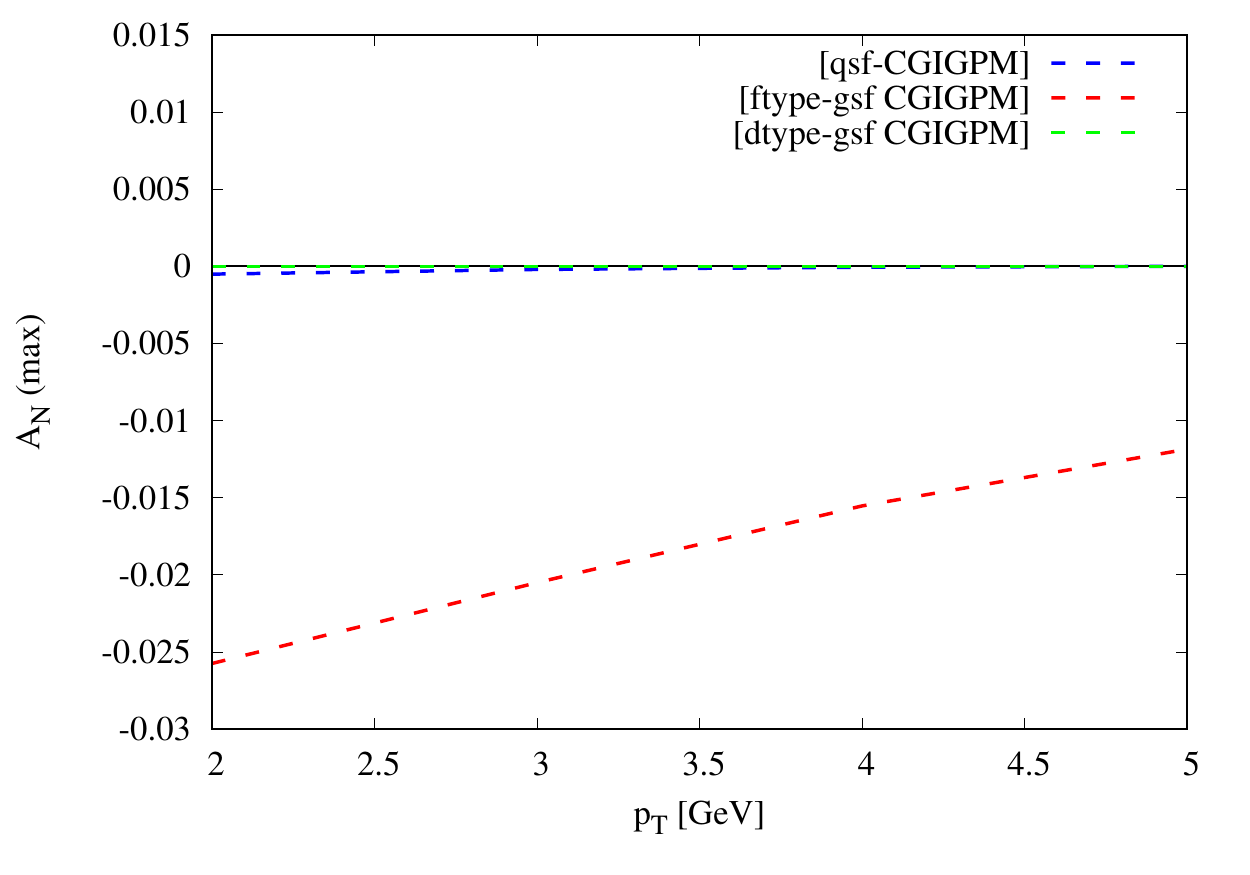}
\caption{ Results for single spin asymmetry in prompt photon production using saturated quark and gluon Sivers function in CGI-GPM. Estimates are given as  a function of $x_{F} $ ( at $ p_{T} = 3 $ GeV, left panel) and $p_{T}$ (at rapidity $\eta =-2$, right panel). Dashed blue line corresponds to the contribution from  QSF, red line corresponds to contribution from \textit{f}-type GSF and green line shows \textit{d}-type GSF contribution.}
\label{fig:an_cgigpm}
\end{center}
\end{figure}

\par In Fig.\ref{fig:an_cgigpm}, we have given estimates of SSA using saturated Sivers functions in the CGI-GPM formalism. As we can see from the plots, saturated QSF contribution is negligible in this case.  This is due to  following two reasons : first,  the quark initiated direct subprocess---which contributes predominantly in the kinematic regions under consideration---does not contribute to  the numerator of SSA in CGI-GPM. Second, in the resolved channel the  quark initiated processes contribute negligibly even  to the unpolarized cross section. In  case of saturated GSF, a \textit{f}-type GSF contribution is negative and is  around 1.5\% to 2.5\% in magnitude, which is  $50\%$ of the  GPM  estimates. The change in sign is  due to the fact that  $ H^{(f)}_{g q(\bar{q}) \rightarrow \gamma q(\bar{q})}=-\frac{1}{2}H^{U}_{g q \rightarrow \gamma q}$.
The reason for vanishing contribution from  \textit{d}-type GSF is that the modified hard parts have opposite signs for quarks and antiquarks $H^{(d)}_{g q \rightarrow \gamma q}=-H^{(d)}_{g \bar{q} \rightarrow \gamma \bar{q}}$. This combined with the fact that the  distribution of quarks and antiquarks is same in the photon leads to zero contribution from \textit{d}-type GSF in this process. The absence of \textit{d}-type GSF contribution to the SSA makes this probe especially useful for extracting information on \textit{f}-type GSF,  which has the dominant contribution to the SSA in CGI-GPM formalism.

\vspace*{2.5cm}
\section{Conclusions}
In this work, we have studied the Sivers asymmetry in prompt photon production in electron proton collisions. We have taken into account both  direct as well as resolved contributions to the production of photons. We find that this probe can be useful for discriminating  between GPM and CGI-GPM as the predictions of asymmetry given by these models are  entirely different. Using GPM, we find, at $x_{F} < -0.2$,  positive asymmetry with sizeable contribution from both the QSF and GSF when using their saturated form. On the other hand, in the case of CGI-GPM,  we find negligible asymmetry due to the saturated QSF while the contribution from \textit{f}-type GSF to the asymmetry is nonzero and negative.
We find that the \textit{d}-type GSF does not contribute to SSA in CGI-GPM.
To conclude, our  initial findings in this study suggest that prompt photon production in low-virtuality electroproduction could be a viable process to investigate the \textit{f}-type GSF in the kinematical region considered here.

\section{ACKNOWLEDGMENTS}
The work of R. M. G. is supported by the Department of Science and
Technology, India under Grant No. SR/S2/JCB-64/2007 under the J.C. Bose Fellowship scheme.  A. M. would like to thank International Centre for Theoretical Physics, Trieste, Italy for their kind hospitality during the completion stage  of this work. S. P. would like to thank Professor V. Ravindran for warm hospitality during his visit to IMSc where the final part of this work was completed.

\section{\label{appdx}APPENDIX: KINEMATICS}
We give here a detailed treatment of partonic kinematics for the direct subprocess. Extension of partonic kinematics for the resolved subprocess should be straight forward by replacing $x_{\gamma_{I}}$ with $x_{\gamma_{I}}x_{b}$. We consider the frame in which the incoming proton and electron are moving along $+Z$ and $-Z$ axis, respectively, and  we fix the scattering plane as the \textit{X-Z} plane.
The four momenta of incoming proton, incoming electron and produced photon are 
\begin{equation}
\begin{split}
P_{P}^{\mu}=\frac{\sqrt{s}}{2}\Bigg( 1,0&,0,1 \Bigg)
, \hspace{1cm}
P_{e}^{\mu}=\frac{\sqrt{s}}{2}\Bigg( 1,0,0,-1 \Bigg)\\
&P_{\gamma}^{\mu}=\frac{\sqrt{s}}{2}\Bigg( E_{\gamma},p_{T},0,p_{L} \Bigg)
\end{split}
\end{equation}
The CM energy of electron-proton system is $s=(P_{P}+P_{e})^{2}$.
We assume that the quasireal photon is collinear to the electron. The quasireal photon and quark four momenta are given by
\begin{equation}
p_{\gamma_{I}}^{\mu}=x_{\gamma_{I}}P_{e}^{\mu}=x_{\gamma_{I}}\frac{\sqrt{s}}{2}\Bigg( 1,0,0,-1 \Bigg)
\end{equation}
\begin{equation}
p_{q}^{\mu}=x_{q}\frac{\sqrt{s}}{2}
\Bigg( 1+\frac{k_{\perp q}^2}{x_{q}^{2}s},
\frac{2k_{\perp q}}{x_{q}\sqrt{s}}\mathrm{cos}\phi_{q},
\frac{2k_{\perp q}}{x_{q}\sqrt{s}}\mathrm{sin}\phi_{q},
1-\frac{k_{\perp q}^2}{x_{q}^{2}s} \Bigg)
\end{equation}
where $x_{q}=p^{+}_{q}/P_{P}^{+}$. 
Using above relations, one can express Mandelstam variables as
\begin{equation}\label{Eq:shat}
\hat{s}=x_{q}x_{\gamma_{I}}s,
\end{equation}
\begin{equation}\label{Eq:that}
\hat{t}=-x_{q}\sqrt{s}p_{T}\bigg[
e^{-y}+
\frac{k_{\perp q}^2}{x_{q}^{2}s}e^{y}-
\frac{2k_{\perp q}}{x_{q}\sqrt{s}}\mathrm{cos}(\phi_{q})
\bigg],
\end{equation}
\begin{equation}\label{Eq:uhat}
\hat{u}=-x_{\gamma_{I}}\sqrt{s}p_{T_{\gamma}}e^{y},
\end{equation}
where $y$ is the rapidity of produced photon.
We impose the following kinematical cuts on Mandelstam variables
\begin{equation}
0 \leq \hat{s} \leq s, \hspace*{1cm}
-\hat{s} \leq \hat{t} \leq 0, \hspace*{1cm}
-\hat{s} \leq \hat{u} \leq 0.
\end{equation}
Furthermore, the inclusion of intrinsic transverse momenta in the kinematics calls for the
following constraints:(i) quark keeps moving along the same direction as proton, $\bf{p}_{q}.\bf{P_{P}}>0$, and (ii) the quark energy is not larger than the proton energy, $E_{q}\leq E_{P}$. This implies the following bound on transverse momentum of quark \cite{DAlesio:2004eso},
\begin{equation}
k_{\perp q}<\sqrt{s}\hspace*{0.125cm} \mbox{min}[x_{q},\sqrt{x_{q}(1-x_{q})}]
\end{equation}

\end{document}